\documentstyle[12pt,fleqn,psfig]{article}
\begin{document}
\newcommand{\tphi}{\tilde{\phi}}
\newcommand{\lton}{\stackrel{\large <}{\sim}}
\newcommand{\gton}{\stackrel{\large >}{\sim}}
\newcommand{\beq}{\begin{equation}}
\newcommand{\eeq}[1]{\label{#1} \end{equation}}
\newcommand{\beqar}{\begin{eqnarray}}
\newcommand{\eeqar}[1]{\label{#1} \end{eqnarray}}
\newcommand{\gfm}{{\rm GeV/Fm}^3}
\newcommand{\half}{{\textstyle \frac{1}{2}}}
\newcommand{\vx}{{\bf x}}
\newcommand{\vq}{{\bf q}}
\newcommand{\vp}{{\bf p}}
\newcommand{\vk}{{\bf k}}
\newcommand{\vK}{{\bf K}}
\newcommand{\vv}{{\bf v}}
\newcommand{\kn}{ $K_0$ }
\newcommand{\knb}{ $\overline{K}_0$ }
\newcommand{\ks}{ $K_s$ }

\begin{flushright}
CU-TP-685\\
\end{flushright}
 \begin{flushleft}
{\Large\bf  Quark Matter 95: Concluding Remarks
 \footnotetext{
*This work was supported by the Director, Office of Energy Research,
Division of Nuclear Physics of the Office of High Energy and
Nuclear Physics of the U.S. Department of Energy under Contract No.
DE-FG02-93ER40764.}}\\[2ex]
Miklos Gyulassy
\\[2ex]
Physics Department,
Columbia University,
New York, N.Y. 10027\\[1ex]
March 1995\\
\end{flushleft}
Abstract: Highlights of Quark Matter 95 are discussed.

\section*{1. The View from Mount RHIC}
This year marked a
 major milestone in the field of high energy nuclear
collisions. Lead beams
\begin{figure}[h]
\hspace{1in}
\psfig{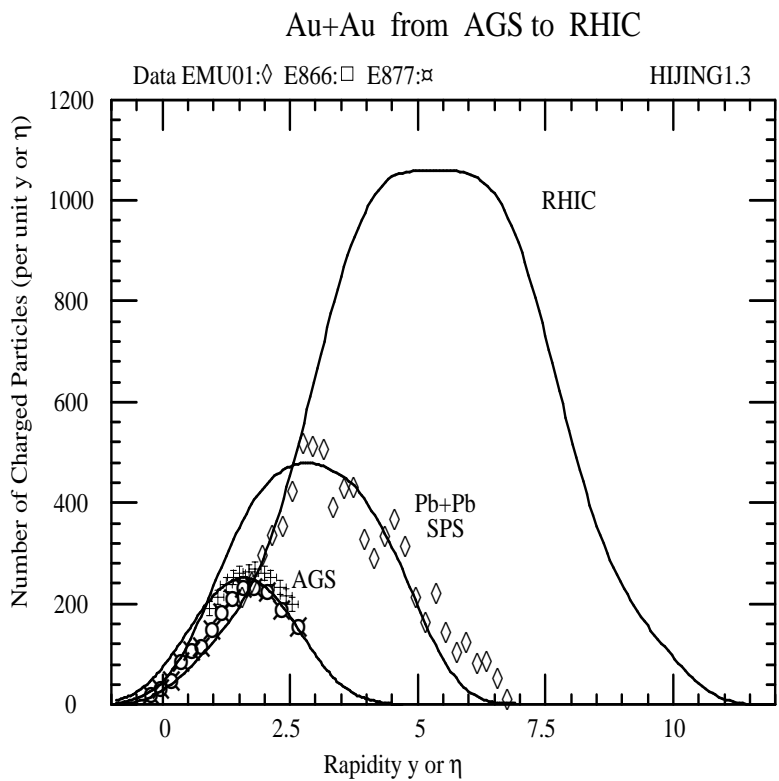}
\caption{The charged particle rapidity density
in central Au+Au (AGS,RHIC) and Pb+Pb (SPS) in the HIJING
model\protect{\cite{hijing}}
compared to pseudo-rapidity, $\eta$,
Emulsion data\protect{\cite{emu}} on the highest multiplicity
event recorded at the SPS to date.
Preliminary E866 data on ($\pi^\pm,\pi^0$)
as well as $dN_{ch}/d\eta$ from E877 at the AGS
are also shown\protect{\cite{videbaek,e877ch}}
}
\end{figure}
were successfully accelerated
to 160 AGeV at the CERN/SPS and the first data
on $Pb+Pb$
interactions were presented at this meeting. At the last quark matter meeting
\cite{qm93} we saw
the first data
on  $Au+Au$ reactions at  BNL/AGS energies (11 AGeV).
Now an entirely new domain of energies has become accessible
with  heavy nuclear beams.
Figure 1
summarizes  where we are now and where
we are going in the next five years.

In the foothills of Mount {\it RHIC}, Figure 1 displays the highest
multiplicity events measured at both the AGS and SPS.
The preliminary pseudo-rapidity
distribution, $dN_{ch}/d\eta$, by the EMU01 collab.\cite{emu} and
from E877\cite{e877ch} and the pion pseudo-rapidity
distribution from
 the E866 multiplicity array\cite{videbaek} are shown.
I remind you that  the  famous
JACEE cosmic ray event\cite{jacee} on $Si+Ag$ at $\sim 5$ ATeV
with $dN_{ch}/dy\approx 200$, which was
used for over a decade to motivate research in this field,
is now overshadowed by $Au+Au$ at the AGS.
The calculated curves are  based  on the HIJING monte carlo
model\cite{hijing} (which combines FRITIOF\cite{fritiof} for
soft beam jet fragmentation and PYTHIA\cite{pythia} for semi-hard
mini jet physics) and are similar to results
obtained with other
models\cite{fritiof,dpm94,venus,rqmd,pcm}
developed in the last few years to calculate
 multiparticle production in nuclear collisions.
We recall  that many of the ideas
now entombed into subroutines  of the above codes
were begot by  the venerable patriarchs in ref.\cite{first}
long before Au was accelerated even in the Bevalac.
The charged particle density  and the transverse energy systematics
reported at this meeting show that
spectacular  $Pb+Pb$ reactions producing $ \sim 1600\;(\pi,p,K)$'s
at 160 AGeV
are, in fact, close $(20\%)$ to  expectations
based on those models. However, as discussed below,
even though the detailed distributions
of hadrons and leptons provide more stringent probes of the dynamics,
it is satisfying that the global characteristics of high energy
nuclear reactions, related to entropy production,
 are under control. This gives us additional confidence
in extrapolations up to the RHIC frontier  that will become accessible
by 1999. Until then there are certainly many interesting trails to explore
in the AGS, SPS foothills.

Before going into further details,
\begin{figure}[h]
\hspace{0.5in}
\psfig{figure=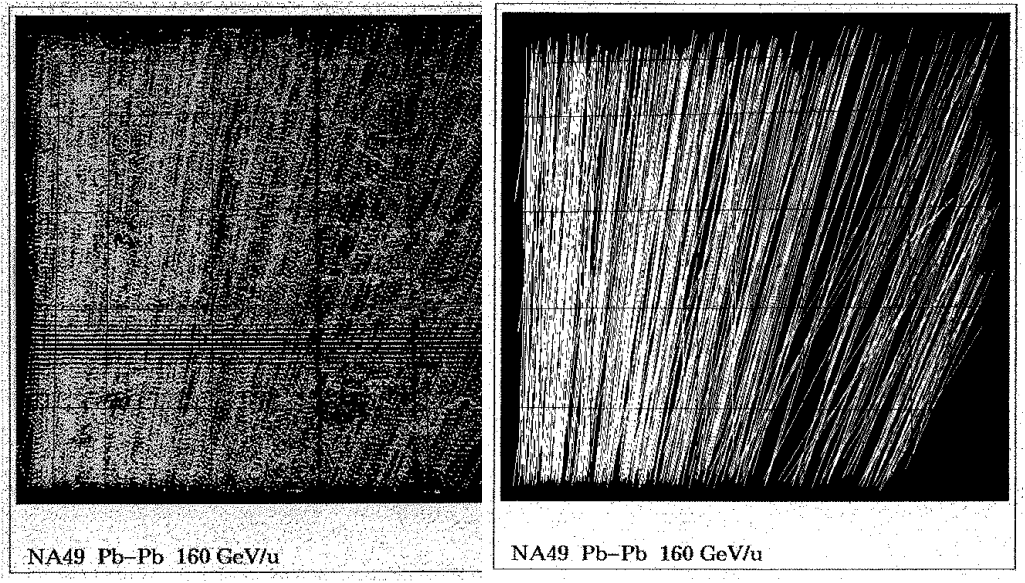,height=3in,width=5.in,angle=-90}
\caption{a) A 160 AGeV  $Pb+Pb$ reaction recorded by one of the NA49 TPC's
at the CERN/SPS.
b) Electronically reconstructed tracks using the 3D information from the TPC.}
\end{figure}
I also want to remark on
the impressive technical
progress made in getting to where we are today.
That progress is well illustrated
in Fig. 2,
 showing
a 2D projection of a Pb+Pb event in one of the new NA49 TPC
detectors\cite{na49}. In the past, such a picture from the streamer
chamber would send shivers down the spine of
poor graduate students  assigned
the task of identifying and tracking the  hundreds
of produced particles. However, the right hand side shows
that because the TPC provides three dimensional data,
tracking such extremely complex events can be done essentially on-line
using the simplest follow-your-nose
tracking algorithms. This provides an existence proof
 for the feasibility of  exclusive measurements at RHIC and LHC
in spite of the enormous multiplicities $\sim 10^4$.
There now exist
adaptive tracking algorithms that could milk
 even multipion correlation functions
out of the jumble of  such events\cite{et}.

Figure 3 compares the spectrum of pion and protons
in $S+S$ at 200 AGeV with  the  PRELIMINARY $Pb+Pb$ reactions
at $160$ AGeV\cite{topor}.
\begin{figure}[h]
\hspace{0.5in}
\psfig{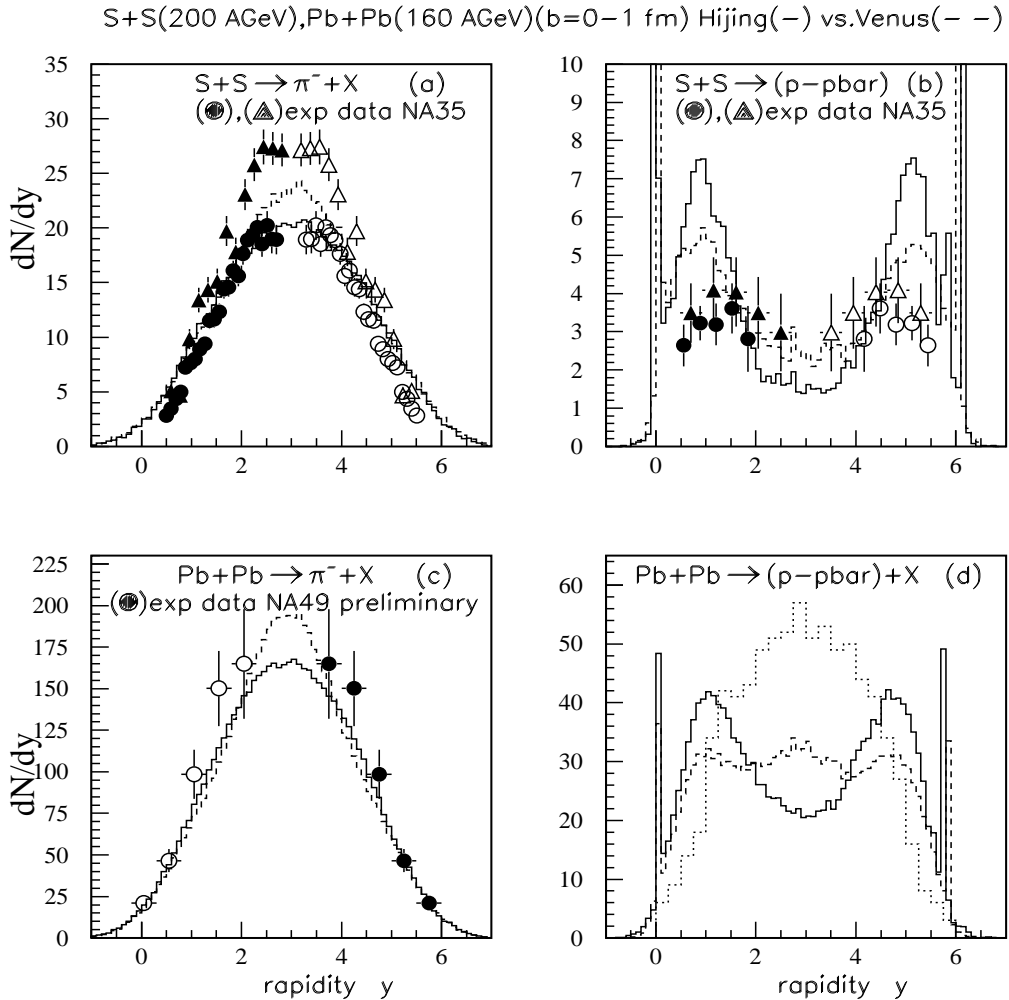}
\caption{Comparison of central $S+S$ at 200 AGeV (a,b) from
NA35\protect{\cite{na35}} to  Preliminary $Pb+Pb\rightarrow \pi^-$
at $160$ AGeV (c,d) from NA49\protect{\cite{na49}}.
Expectations based on
HIJING\protect{\cite{hijing}}, VENUS\protect{\cite{venus}},
and RQMD\protect{\cite{rqmd}}
models are depicted as solid,dashed, and dotted
histograms, respectively\protect{\cite{topor}}.
}
\end{figure}
The various data sets from NA35 for
$S+S\rightarrow \pi^-$
correspond to different centrality triggers,
with the higher one corresponding to a more severe veto trigger cut.
We see that the negative pion rapidity densities
are well accounted for by both
 HIJING\cite{hijing}
 and VENUS\cite{venus}, but the flat valence proton distribution
in $S+S$ is only reproduced by VENUS. Recall  that
VENUS, like RQMD, includes a model of final state interactions
in dense matter as well as a color rope effect, called double strings.

The main experimental unknown as yet  is the fate
of the baryons in the $Pb+Pb$ reaction illustrated
in Fig.3d. As discussed by Sorge
in this meeting, the  RQMD model\cite{rqmd,rqmd_stop}
predicts a much higher degree of
baryon stopping at midrapidity
 than VENUS,
and HIJING as well as   FRITIOF predicts a hole at midrapity.
These very large differences between predicted valence proton spectra,
in contrast to the inclusive pion spectra in parts a,c,
are due to the very different  dynamical assumptions
associated with nuclear stopping power.
Baryon stopping is limited in the FRITIOF type models like
 HIJING
by the assumption that
a diquark propagates as a hard nugget unscathed through a nucleus
and fragments into a baryon and mesons
only outside the nucleus due to time dilation\cite{feinberg}.
In RQMD the large nuclear stopping power is due
to  the  assumption that multiple collisions of diquarks in nuclei
can be treated incoherently, neglecting formation time physics.
In VENUS, extending the Dual Parton phenomenology\cite{dpm94},
the diquarks are allowed to disintegrate and form  double strings,
which make it possible to shift the valence baryon number
further
away from the fragmentation regions. The VENUS model parameters
are  tuned to reproduce available $p+A\rightarrow p+X$ data
and comes closest to the expectations
for baryon stopping emerging from  earlier studies\cite{date}.
My bet is that this is where the data will land unless
molten lead has a big surprise in store for us.

The final experimental resolution of nuclear stopping power
problem must await the  next quark matter meeting.
This problem is of fundamental interest because
it is related to how high the baryon densities may get in such reactions.
Are we still in the baryon stopping regime at the SPS, as is the case
at the AGS (see S. Margetis\cite{qm95}), or
are reactions at SPS making the transition to the low baryon density regime
expected at RHIC?
In the first case, the dynamics of shock formation and Landau hydrodynamics
may be relevant. In the second case, the Bjorken
longitudinal expansion and inside-outside cascade dynamics
is more relevant.

\section*{2. The Flow of Gold}

The discovery of collective sidewards flow in Au+Au at the AGS\cite{zhang}
is a major highlight this year.
It shows the persistence of collective flow
phenomena all the way up to  AGS energies. This phenomena
was first discovered\cite{ball} at Bevalac energies in 1984
when   heavy nuclear beams first became available.
It is of fundamental importance because it provides a direct
probe of  the equation of state at extremely high densities\cite{stocker}.

E877\cite{zhang} found as shown in Figure 4,
that the
distribution of the normalized transverse energy dipole
moments
$$v_1\propto \left[(\sum_i E_{\perp i}\cos(\phi_i))^2 +
(\sum_i E_{\perp i}\sin(\phi_i))^2 \right]^{1/2}$$
in $Au+Au$ is systematically shifted toward
 finite values for more central collisions.
The sum above is over calorimeter modules
sensitive to  only forward of $y_{cm}$ fragments
with different azimuthal angles $0\le \phi_i\le 2\pi$.
This variable provides  one
of the  measures of the collective transverse flow pattern
of the system.

The quantitative evaluation of such flow data requires the
use of elaborate transport codes. In the lower energy domain,
\begin{figure}[h]
\hspace{1in}
\psfig{figure=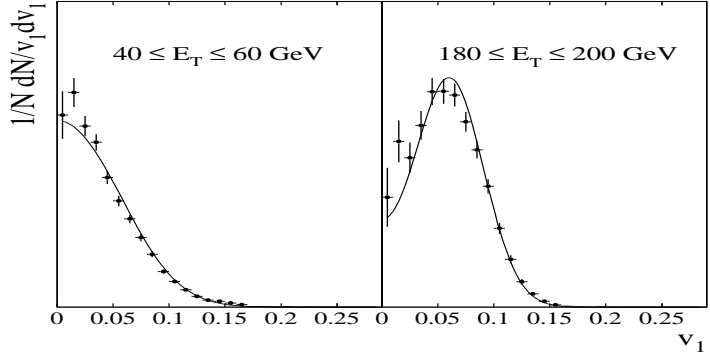,height=2in,width=4in,angle=0}
\caption{E877\protect{\cite{zhang}} data showing evidence
of  transverse collective flow
in $Au+Au$ reactions at 10 AGeV.
Left side corresponds to peripheral collisions  and right side to
mid-impact parameter collisions.}
\end{figure}

\begin{figure}[h]
\hspace{1cm}
\psfig{figure=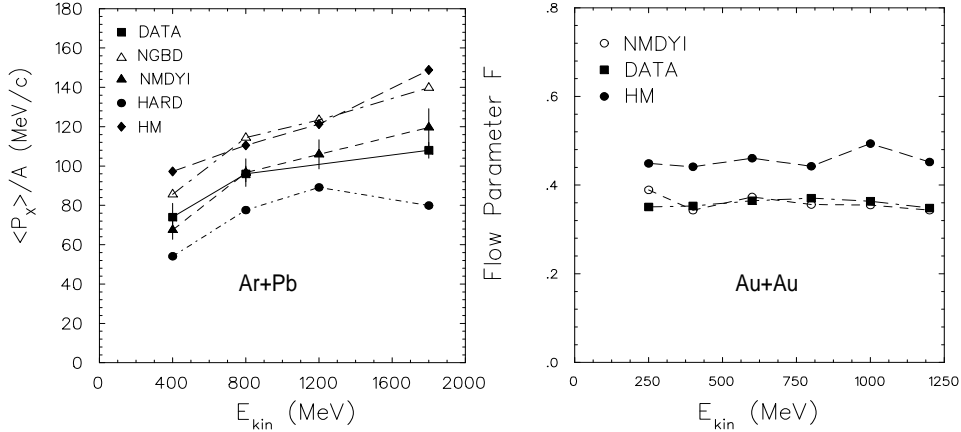,height=2.8in,width=5in,angle=-90}
 \vspace{-1.5cm}
\large
\caption{ a) Mean in plane transverse momentum $(y_{cm}>0)$
in Ar+Pb \protect{\cite{beavis}}
Flow parameter $F=d\langle p_x\rangle/dy|_{y=y_{cm}}$
in Au+Au \protect{\cite{eos}}
}
\end{figure}
it took a decade of work by many groups
to achieve what is now a truly impressive  degree of convergence.
This is  because sufficiently high precision
and detailed triple differential cross sections
have only become available with  technical
developments such as the EOS/TPC\cite{eos}
and major improvements in the theoretical transport
tools were required\cite{daniel}.
Figure 5 shows the results of  the most recent analysis of ref.\cite{gale}
of the mean in-plane transverse collective momentum per nucleon
in $Ar+Pb$ and $Au+Au$ reactions from $0.2-1.2$ AGeV.
What is so remarkable about these results is that all the flow data
on asymmetric as well as symmetric nuclear collisions are reproduced
by  a BUU transport model
taking a momentum and density dependent optical potential (NMDYI)
that is {\em consistent} with the Urbana UV14+UVII force.
That force  is known\cite{wiringa} to provide a good
description of light nuclei and bulk nuclear
matter properties and has a compressibility $K=210$ MeV consistent
with nuclear breathing modes.
This is the first time that nuclear collective flow
in nuclear collisions can be explained quantitatively in terms
of well known nuclear interactions. Actually, the flow data
is providing new information on aspects of the nuclear interactions
not tested by ordinary nuclear properties near saturation.
However, a consistent picture is beginning to emerge that
links not only low energy nuclear phenomena to collective flow
in low and intermediate energy nuclear collisions, but also
to the types of equations of state used in calculating the bounce in supernovae
and the structure of neutron stars.

Figure 5 establishes therefore a very important
fixed point from which explorations
deeper into the high baryon density domain can be based.
At Bevalac and now GSI/SIS energies the compressions
are modest $\rho_B < 4 \rho_0$, and it may not be too surprising that
conventional nuclear theory works so well.
However, at the AGS energies, all estimates indicate that
baryon densities up to $10\rho_0$ are generated.
In that case, a breakdown of extrapolations of conventional
nuclear physics may occur since the baryons and mesons
are squeezed on top of each other, possibly melting into a quark plasma.
 In order to identify any new physics  associated with
the expected QCD phase transition, of course much more work will
be needed both experimentally and theoretically.
On the experimental side, it will be necessary to
measure detailed triple differential cross sections
of identified fragments and especially to study the beam energy dependence
of the flow phenomena (H. St\"ocker\cite{qm95}).
 At present the first generation
of  cascade models
like ARC\cite{arc,arcflow} apparently reproduce the trends
seen in Figure 4  (see talk by
Y. Pang\cite{qm95}), but only at the price of introducing assumptions
about  hard core,
classical repulsive scatterings as in earlier models for flow
at Bevalac/SIS energies\cite{mgstocker}.

As shown at the last quark matter\cite{bravina},
flow phenomena at AGS energies are  of  interest
because they are  sensitive to possible
 softening of the equation of state
across the hadron to quark-gluon plasma transition.
Depending on the nature of the QGP transition at high baryon density,
the  crossover energy \cite{ags_hydro}
between the hadronic and quark phases may occur anywhere
 between $E_{lab}=2-10$ AGeV. Thus, the current $Au+Au$ AGS
experiments at 10 AGeV may have overshot the transition region!
Indeed, a very speculative interpretation\cite{pursun} of the
relatively small flow observed in the E877 experiment in Figure 4
is that this may   be already the direct consequence of the soft
 equation of state in the QCD plasma phase.
The most striking prediction is that if this is the case,
then the degree of flow should increase with {\em decreasing}
lab energy below $10$ AGeV!
I feel that this is certainly one of the most exciting directions
to pursue experimentally at the AGS in the next several years
along with the search for exotic multistrange objects (S. Kumar\cite{qm95}).
Such phenomena if observed would be very difficult to imitate with cascade
models. If, on the other hand,
 the flow is shown to remain independent of energy and  continues
along the same flat curve as in Fig. 5b, then
an abrupt QGP transition at high $\rho_B$
could be ruled out. In any case, the EOS/TPC used in ref.\cite{eos}
would be an ideal tool for such further studies at the AGS.

\section*{3. Gold is Strange, But $p+S$ is Weirder}

An important diagnostic of dense matter formed in  nuclear collisions
is strangeness\cite{raf1}. The E802 team
(B. Cole\cite{qm95}, Z. Chen\cite{qm95}) has mapped out
with most precision the systematics of
strangeness  enhancement at the AGS\cite{cole}.
In Figure 6, the old together with the new data on $Au+Au$
\begin{figure}[h]
\hspace{1in}
\psfig{figure=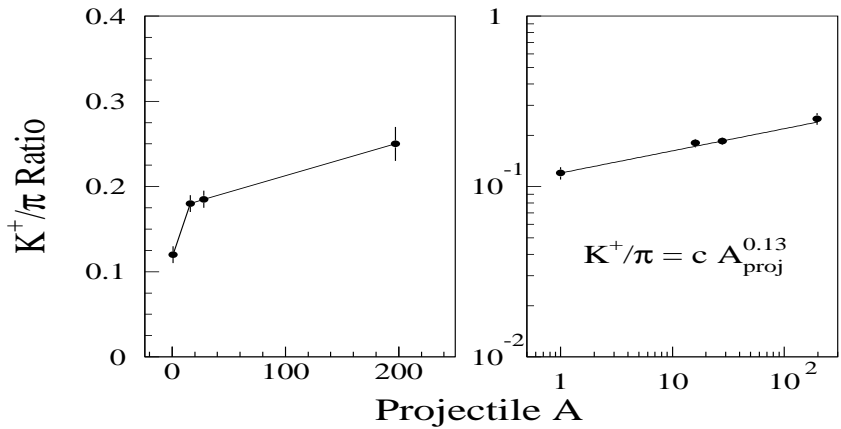,height=2.5in,width=5in,angle=0}
\caption{E802 data on $K^+/\pi$ enhancement in $A+Au$ reactions at 11
AGeV\protect{\cite{cole}. The $Au+Au$ point is  PRELIMINARY.}
}
\end{figure}
are shown. Most conspicuous is the
apparent saturation effect of the $K/\pi$ ratio
as the size of the projectile nucleus increase. Also
the very weak power dependence of $K/\pi\sim A^{0.13}_{proj}$
\begin{figure}[h]
\hspace{1in}
\psfig{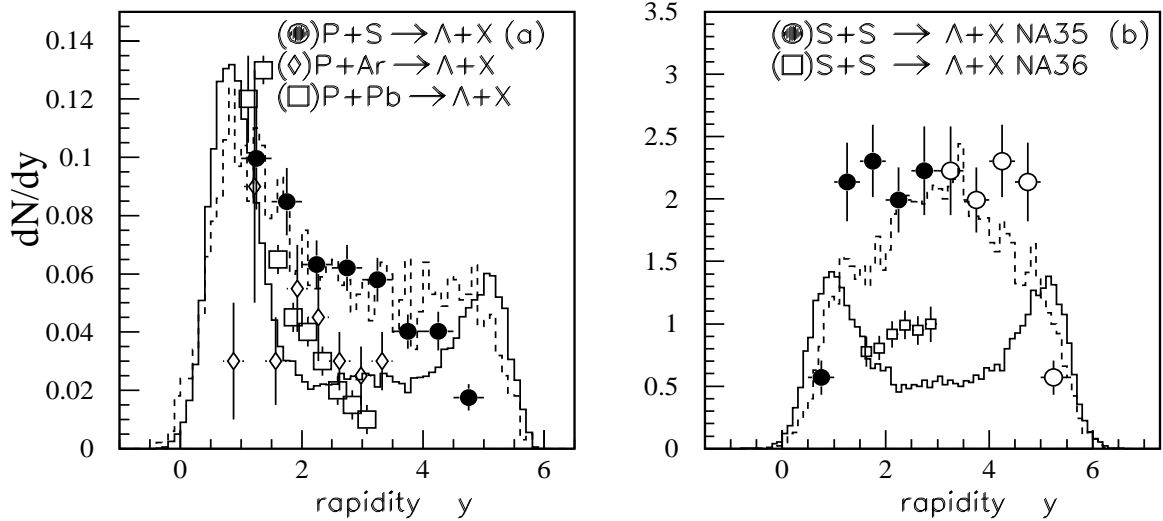}
\caption{$\Lambda^0$ production in $p+S$ and $S+S$ at 200 AGeV
 from NA35\protect{\cite{na35_lam}} and NA36\protect{\cite{na36}}
compared to HIJING and VENUS results\protect{\cite{topor}}
}
\end{figure}
from $A_{proj}=1$ to 197 is remarkable.
These results suggest that strangeness production
in $A+A$  smoothly extrapolates  from $pp$ to $pA$
to  $AA$. There is obviously no threshold effect indicating the onset
of any equilibrium source of strangness.

This conclusion is brought into even clearer focus  in
Figure 7 comparing $\Lambda^0$  production in $p+S$
and $S+S$ reactions at SPS energies\cite{topor}.
Again HIJING and VENUS calculations are contrasted.
Data from NA35\cite{na35_lam} (M. Gazdizcki\cite{qm95})
and NA36\cite{na36} (E. Judd\cite{qm95}) are also compared.
As in the last quark matter conference, there is a  significant
discrepancy between the two data sets.
The weirdest result is the NA35 $pS$ data
that exceeds the NA36 $p+Pb$ data and lies a factor of
two above the HIJING results\cite{topor}.
In ref.\cite{topor}, it was shown on the other hand
that HIJING reproduces well the observed
$\Lambda^0$ production in $pp$.
There also appears a large difference between VENUS
and HIJING  results.
This is especially remarkable given that minimum biased $pS$
is the most boring extension of $pp$ collisions imaginable!
On the average, in $pS$, the projectile nucleon interacts with only two target
nucleons. Nothing could be further from the thermodynamic limit,
except of course $pp$. How could extrapolations from
$pp$ breakdown so quickly
in $p+2p$. The trick invoked in VENUS, as well as in RQMD\cite{rqmd_rope},
is to invoke the color rope idea\cite{biro}.
Overlapping strings even in $p+A$ from different target
nucleons may fuse into one with a higher string tension.
Pair production in such enhanced color fields
can  easily produce hoards of extra strange and even charm quarks.

In Fig 7b, midrapidity $\Lambda$'s are enhanced by another
factor of two in $S+S$, at least in the NA35 data.
For heavier targets there is a saturation effect as in Figure 6a.
Therefore all the  enhancement of strange baryons can be traced back to
the rapid increase of strangeness in the {\em non-equilibrium}
dynamics of $p+p+p$ and $p+p+p+p$. It is therefore simply not relevant,
in my
opinion, to apply thermal fireball models to explain strangeness
enhancement. It has to do instead with interesting new dynamical effects,
possibly along the rope ideas,  and little to do with the QGP transition.
The problem is not to reproduce the ratios of integrated yields,
but to explain quantitatively the distributions in the weird $p+A$ systems.

\section*{4. The Shine of Gold}

While direct $\gamma$'s from  light ion reactions
continue to elude  WA80 (T. Awes\cite{qm95}),
 the very dimness of
the emitted light delighted
Srivastava\cite{qm95}, who claimed that this was the smoke-less gun that
proved the QGP transition.
NA45 presented new data\cite{ceres_na45}
(P. Wurm\cite{qm95}, I. Tserruya\cite{qm95}), on the other hand,
revealing an excess of dilepton pairs in the mass region $2 m_\pi<m<1.5$ GeV
for $S+Au$. A similar effect was reported by the HELIOS collaboration
(M. Masera\cite{qm95}),
which also showed  an excess in the intermediate
mass range $m
\sim 1.5\pm 2.5 $ GeV as compared to $p+A$.
In $p+A$ the observed pairs light mass pairs were
well  accounted for by $\eta$ and $\omega$ Dalitz
decays. However, extrapolations of those
backgrounds to $S+Au$ appears to under-estimate
the observed yield by a factor $5\pm0.7\pm2$.
This may indicate new physics or the onset
of final state processes such as $\pi\pi\rightarrow e^+e^-$.

Another rare probe that showed hints of unusual behavior
is the mass specturm of the $\phi$ meson. Last meeting, Y.  Wang
reported the first
successful measument of $\phi$ spectra in $Si+Au$ at the AGS.
This meeting (see Y. Wang\cite{qm95}) hints for a tiny few MeV shift
were presented. This is much smaller than the spectacular
100 MeV shifts predicted  by Ko and Asakawa\cite{asakawa}
as a signature of the chiral restoration transition  in the dilepton channel,
but their is not much phase space  in the $KK$ decay channel.
A  few MeV shift is consistent with the uncertainty principle only
if the system lived $\sim 100$ fm! In any case,
it could be an interesting interference or final state
effect  and we should keep an eye
on this problem.

The most well established and hotly debated
rare probe since quark matter 1987
is the $J/\psi$ and now $\psi^\prime/\psi$
suppression. The data from NA38 (see S. Ramos\cite{qm95}) are quite convincing
of a true nuclear suppression effect that must involve the formation
of a very dense comoving system.
The debate focuses on  whether there is any evidence
for a threshold effect or,  as is in the case of the strangeness enhancement,
this  phenomenon extrapolates smoothly back to $pA$ and $pp$.
Last quark matter conference,
 Gavin\cite{gavin} presented a strong case that the
suppression
of hidden charm smoothly extrapolates down from $S+U$ back to $pp$
and can be understood quantitatively if one assumes a comover
density $\sim 5\rho_0\approx 0.8 $ fm$^{-3}$
and a dissociation  cross section of a few mb.
This meeting D. Kharzeev (see these proceedings\cite{qm95})
challenged those finding with a new estimate based on
heavy quark mass limit of QCD arguing that  the
$J/\psi$ dissociation cross section should be  much smaller.
However, the burden on theory  is then again shifted
to explain the suppression
in the manifestly {\em non-equilibrium} conditions of $p+A$ and light
ion induced reactions. It will be  interesting to see
if there is enhanced suppression in upcoming $Pb+Pb$ measurements
beyond that predicted by the comover model\cite{gavin}.

\section*{5. The Charm of Gold}

The possibility that at RHIC energies nuclear collisions
will become even more charming
\begin{figure}[h]
\psfig{figure=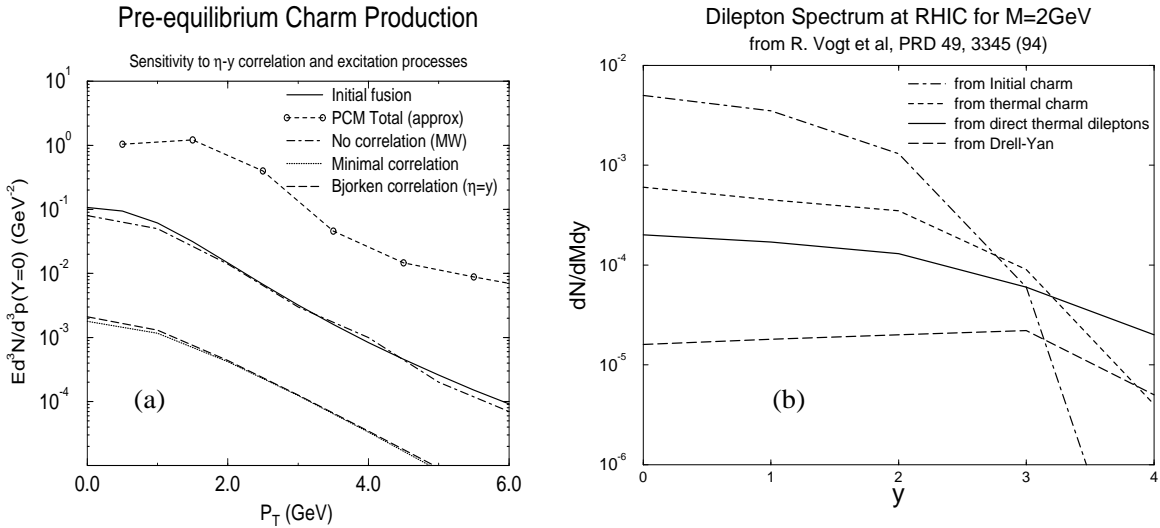,height=2.7in,width=6in,angle=-180}
\caption{(a) Comparison of open charm transverse momentum
distribution produced in $Au+Au$ at 200 AGeV. Predictions
of ref. \protect{\cite{pcm,MW,lin}} are shown.
(b) The dilepton rapidity spectrum for $M=2$ GeV
is dominated by open charm decay, from ref.\protect{\cite{vogt}}
}
\end{figure}
 was emphasized by K. Geiger\cite{geiger_c} and
R. Vogt\cite{vogt} at the last meeting. In Figure 8a, the initial gluon fusion
rate into $c\bar{c}$ pairs is compared to three models.
The highest curve is the prediction of the PCM model\cite{geiger_c}
that predicts a factor of ten enhancement, the Muller-Wang fireball
model\cite{MW} (MW)
that predicts comparable yield from the pre-equilibrium
stage after mini-jet formation,
 and the recent calculation in ref.\cite{lin} (see Z. Lin\cite{qm95}, these
proceedings)
that predicts a factor ten less than initial fusion.
The order of magnitude differences
between these theoretical predictions is due to
different assumptions for the intrinsic charm component and
phase space correlations in the plasma. Most of the charm of PCM comes from
an enormous intrinsic charm component assumed in the nucleon structure
function (GRV), which now appears to be ruled out\cite{lin}.
The comparable charm in \cite{MW} arises from the assumption that
the plasma fireball decouples to avoid the Bjorken longitudinal
expansion. In \cite{lin} both the pre-equilibrium and equilibrium charm
production is found to
be suppressed even if the ideal $y=\eta$ Bjorken
phase space correlations are smeared by the Bjorken cloud.
The conclusion is that most of the open charm expected comes from the initial
gluon fusion stage.

A similar conclusion was found  by Vogt et al\cite{vogt}
shown in Figure 8b. Here the contributions to the
 observable dilepton spectrum from different sources are shown.
Clearly the dilepton spectrum in the $M=2$ GeV range should be an ideal probe
of the initial gluon fusion into charm pairs. As such it is an ideal
probe of the nuclear gluon structure function.
Thus the dilepton measurements in this mass range
will provide essential information on gluon shadowing and anti-shadowing
(see talk of K. Eskola\cite{qm95}) so essential for the mini-jet physics that
makes mount RHIC tower in Figure 1 so much higher than
the current SPS range. PHENIX will be the detector of choice for
this observable.

\section*{6. The Color of Gold}

T. D. Lee\cite{qm95}  reminded us that the main reason we
are in this business of inverse alchemy (transforming Au into
a colorful, strange and charming topless-bottomless  quark-gluon plasma),
 is to wreak havoc on the
non-perturbative QCD vacuum.
The Au  beams are the bulldozers
with which we hope to sweep away
the vacuum  condensates over a  large space-time volume.
The ultra-dense matter formed in their wake, at least at RHIC
energies and above,  should be most economically
described in terms the fundamental   quark and gluon
degrees of freedom rather the hoards of resonances
in the particle data book or the subroutines of event generators.
As discussed by  K. Eskola\cite{qm95} and X.N. Wang\cite{qm95},
high energy nuclear collisions
provide the unique tool to
 probe  experimentally the structure of the physical
vacuum by  heating it to  at least 10 GeV/fm$^3$
in the form of  $\sim 1000$ mini-jets.
At present AGS and SPS energies, the energy densities are
significantly smaller but
perhaps sufficient to penetrate through the intermediate mixed
phase.
The precise nature of the QCD transition remains
unclear. In Figure 9, the state the art\cite{gott} from  the lattice
QCD  as discussed by F. Karsch\cite{qm95} is shown.
This is similar to many previous calculations but perhaps
closer to the continuum limit, and also the temperature
scale is fixed using the nonperturbative beta function
and the $\rho$ mass measured on the lattice.
It shows that the transition is perhaps continuous but still
confined to a rather narrow temperature range.
The perturbative Stefan Bolztmann domain may hold approximately already at $
T> 2 T_c\approx 300$ MeV, which is easily reached at RHIC.
Near the mixed phase region, significant deviations from ideal behavior
is  expected and maybe that is why hadronic
tranport models are working so well at present
AGS and SPS energies.

T. D. Lee\cite{qm95} also presented  a new theory\cite{tdlee2}
involving  a non-compact formulation
of lattice QCD that removes spurious  fermion modes while retaining
 computation advantages of a finite lattice.
A complete set of Bloch wavefunctions was proposed
 in which to expand the wavefunction of
the QCD vacuum. The next step requires
 a clever
choice of a trial wavefunction with the goal of
computing systematic corrections perturbatively
 including higher bands.
\begin{figure}[h]
\hspace{1in}
\psfig{figure=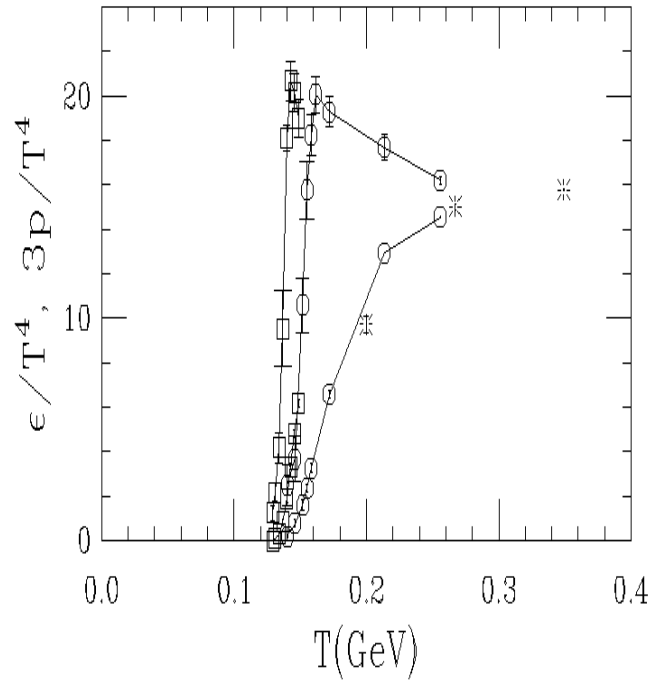,height=4in,width=4in,angle=-90}
\caption{The energy density and pressure in units ot $T^4$ as a function of
temperature for $N_f=2$ QCD on a $12^3\times 4$ lattice
from \protect{\cite{gott}}. The  squares and circles are for quark masses
$m_q/T=1/4,1/10$ respectively.
}
\end{figure}

On the Disordered Chiral Condensate front introduced last meeting by
Wilczek\cite{qm93},
we heard several progress reports.
 Gavin\cite{qm95} discussed its formation and proposed
several observables that may help look for them. In particular
there may be a  low $p_T$ enhancement above the huge incoherent background.
However,
Asakawa\cite{qm95} showed that the
 so called annealing scenario does not favor DCC
formation using a simulation that includes both transverse
and longitudinal expansion.  The original rapid quench
scenario, which seems to allow for the growth of DCC crystals,
requires on the other hand a miraculous inverse Baked-Alaska
scenario proposed by Bjorken, whereby
the hot plasma ejects rapidly  its enourmous entropy
and leaves a cold chunk of disoriented vacuum
behind. This is a looong shot, but worth searching for.

A major new development reported at this meeting
by Venugopalan\cite{mclerran}
was  a theory for the gluon and quark structure functions
for very heavy nuclei.
All parton cascade models, such as HIJING, VENUS and PCM,
must  make assumptions on how the structure
functions of nuclei may differ from $A^1$ times that of nucleons.
Shadowing and anti-shadowing effects can significantly modify
the initial conditions (see  Wang and Eskola\cite{qm95}).
The initial conditions of course  control
 the  final observables.
For example,
Mount {\it RHIC} in Figure 1 could be 2-4 times as high,
as in the PCM model\cite{pcm}. This translates into
{\em macroscopic}
differences in the final transverse energy
 corresponding to several {\em ergs }
per unit rapidity!! This uncertainty originates
from the poorly known
early evolution of the mini-jet plasma.
By developing a systematic treatment of the origin of the non-abelian
Weizsacker-Williams fields around nuclei, it should  be possible
to compute the early evolution of the color fields more reliably.
This method should enable us to estimate
 the height of Mount {\it RHIC} better in the near future\cite{int}.\\[2ex]

Acknowledgements: I am especially grateful
to M. Asakawa, Z. Lin, D. Rischke,  V. Topor Pop, and B. Zhang
for assitance and extensive discussions.
I thank K. Werner for permission to use VENUS and H. Sorge
for providing the proton rapidity density from RQMD.
I apologize to the many speakers  whose interesting results I have not
had time to discuss
(or  misrepresented) in this condensed review.
 \\[2ex]


\begin{thebibliography}{10}
\bibitem{qm93} Quark Matter '93, eds. E. Stenlund, H.A. Gustavson,
A. Oskarsson, and I. Otterlund, Nucl. Phys. A566 (1994) 1.

\bibitem{emu} E. Stenlund, these proceedings;
EMU01 Collaboration. (J.I. Nystrand, et al.). 1994.
Nucl. Phys. A566 (1994) 419c-422c;
(M.I. Adamovich, et al.), Phys. Lett. B322 (1994) 166-170.

\bibitem{videbaek} F. Videbaek (E866 Collab) these proceedings.

\bibitem{e877ch} J. Stachel, private communication.

\bibitem{jacee} JACEE collab., T.H. Burnett et al., Phys. Rev. Lett.
50 (1983) 2062.


\bibitem{hijing} X.N. Wang and M. Gyulassy,  PRD44 (1991)
3501;  PRD45 (1992) 844; Comp. Phys. Comm. 83 (1994) 307.

\bibitem{fritiof}B.Andersson, et al, Nucl.Phys.{\bf B281},289(1987);
Comp.Phys.Commun.{\bf 43},387(1987)

\bibitem{pythia}T.Sjostrand, Comp.Phys.Comm.{\bf 82},74(1994).

\bibitem{dpm94} A.Capella,U.Sukhatme,C.~I.~Tan and
J.~Tran ~Thanh ~Van ,Phys.~Rep.{\bf 236},225(1994)

\bibitem{venus} K.~Werner,
Phys.Rep.{\bf 232},87(1993).

\bibitem{rqmd} H. Sorge,H. Stocker and W. Greiner,
Nucl.Phys.{\bf A498}, 567c(1989); Nucl.Phys.{\bf A566}, 633c(1994).

\bibitem{pcm} K.Geiger and B.Mueller,Nucl.Phys.{\bf B369},600(1992);
 Phys.Rev.{\bf D47},133(1993);
Preprint CERN, CERN - TH -7313 (1994).

\bibitem{first} S.J. Brodsky, A. Bialas, A. Capella, in 1st Workshop
on Ultra-Relativistic Nuclear Collisions, LBL-8957, UC-34c,
CONF-7905107 (May 1979).

\bibitem{na49} NA49 Collaboration, I thank J. Harris and R. Stock
for these pictures.

\bibitem{et} M. Gyulassy and M. Harlander,  Comp. Phys. Comm. 66 (1991) 31;
Nucl. Inst. Meth. A316 (1992) 238.

\bibitem{topor} V. Topor Pop, et al, Columbia preprint CU-TP-676;
and to be published.

\bibitem{feinberg} E.L.Feinberg, Phys. Rep. 5 (1972) 237;
J. Koplik, A.H. Mueller, PRD 12 (1975) 3638.

\bibitem{rqmd_stop} A. von Keitz, et al, Phys.Lett.B263:353-358,1991.

\bibitem{date} W. Busza, R. Ledoux, Ann.Rev.Nucl.Part.Sci.,
ed. J.D. Jackson, Vol. 38 (1989) 119;
 S. Date, M. Gyulassy and H. Sumiyoshi,
   Phys. Rev. D32, 619 (1985).


\bibitem{na35} J. B\"achler, et al (NA35 Collab.) Phys. Rev. Lett.
72 (1994) 1419
\bibitem{na35_lam} T.Alber et.al,
Z.fur Physik {\bf C64},195(1994)

\bibitem{qm95} Quark Matter 95, these proceedings.

\bibitem{arc} S.H. Kahana, T.J. Schlagel, Y. Pang,
Nucl. Phys. A566 (1994) 465c;
Y. Pang, T.J. Schlagel and S. Kahana, Phys. Rev. Lett.
{\bf 68}, 2743 (1992).

\bibitem{zhang} J. Barrette, et al (E877 Collab.),
 Phys. Rev. Lett. 73 (1994) 2532.

\bibitem{ball} H.H. Gutbrod, A. M. Poskanzer, and H.G. Ritter,
Rep. Prog. Phys. 52 (1989) 1267.

\bibitem{stocker} H. St\"ocker and W. Greiner, Phys. Rep. 137 (1986) 277.

\bibitem{beavis} D. Beavis et al, PRC 45 (1992) 299.

\bibitem{eos} M.D. Partlan et al (EOS Collab.), LBL-36280 (1994).

\bibitem{daniel} Q. Pan, P. Danielewicz, Phys. rev. Lett. 70 (1993) 2062.

\bibitem{gale} J. Zhang, S. Das Gupta, C. Gale, Phys. Rev. C50 (1994) 1617.

\bibitem{wiringa} R.B. Wiringa, Phys.ReV.C38 (1988) 2967.


\bibitem{bravina} L.V. Bravina, et al, Nucl.Phys. A566 (94) 461c.

\bibitem{pursun} Y. P\"urs\"un, Diploma  Thesis (1994)
unpublished.

\bibitem{raf1}P.Koch, B.Muller and J.Rafelski
Phys.Rep.{\bf C142} 167 (1986).

\bibitem{rqmd_rope} H. Sorge,
M. Berenguer, H. Stocker, W. Greiner, Phys.Lett.B289:6-11,1992.



\bibitem{arcflow}D. E. Kahana, D. Keane, Y. Pang, Tom
Schlagel, S. Wang,
NUCLTH-9405017 (1994).

\bibitem{ags_hydro} D.H. Rischke, H. St\"ocker, W. Greiner,
J.Phys.G14:191,1988; Phys.Rev.D41:111,1990;
N. K. Glendenning, Nucl.Phys.A512:737,1990.

\bibitem{mgstocker}  M. Gyulassy, K.A Frankel, H. St\"ocker
   Phys. Lett. 110B, 185 (1982).


\bibitem{cole} B. A. Cole et al, E802 Collab,
Proc. 5th Int. Conf. on Intersection of Part. Nucl. Phys. (1994).

\bibitem{na36} D. Greiner, et al, (NA36) preprint  LBL-36882 (1995);
 E.Andersen et.al.,
Nucl.Phys. {\bf A566},487c(1994).

\bibitem{biro} T.S. Biro, H.B. Nielsen, J. Knoll,
Nucl. Phys. {\bf B245},449(1984);
M. Gyulassy, A. Iwazaki, Phys. Lett. 165B (1985) 157.

\bibitem{ceres_na45} CERES/NA45 Collab. (G. Agakichiev, et al.)
CERN-PPE/95-26 (1995).

\bibitem{helios} HELIOS/3 Collab. (M.A. Mazzoni, etal).
Nucl. Phys. A566 (1994) 95c-102c.


\bibitem{asakawa} M. Asakawa, C.M. Ko (Texas A-M). 1994.
Nucl. Phys. A572 (1994) 732-748.

\bibitem{na38}  NA38 Collab. (B. Ronceux, et al.),
 Phys.Lett.B345 (1995) 617; Phys. Lett. B255 (1991) 459-465.

\bibitem{gavin} S. Gavin,Nucl. Phys. A566 (1994) 383c; S. Gavin and R. Vogt,
Nucl.Phys.B345 (1990) 104.



\bibitem{geiger_c} K. Geiger, Phys. Rev. D48, 4129 (1993).

\bibitem{MW}
B. M\"{u}ller and X. N. Wang, Phys. Rev. Lett. 68, 2437 (1992).

\bibitem{lin} Z. Lin and M. Gyulassy, Columbia Preprint CU-TP 638 (1994),
PRC in press.

\bibitem{vogt} R. Vogt, B.V. Jacak, P.L. McGaughey , P.V.
Ruuskanen, Phys.Rev.D49 (1994) 3345.

\bibitem{SX}
E. Shuryak and L. Xiong, Phys. Rev. Lett. 70, 2241 (1993).


\bibitem{tdlee} T. D. Lee, CU-TP-678 (1995), these proceedings.


\bibitem{gott} T. Blum, L. K\"ark\"ainer, D. Toussaint,
S. Gottlieb, Preprint AZPH-TH/94-22, IUHET-284 (1994).

\bibitem{tdlee2} R. Friedberg, T.D. Lee, Y. Pang, H.C. Ren,
J. Math. Phys. 35 (1994) 5600;
CU-TP-662 (1995) preprint.

\bibitem{mclerran}  L. McLerran, R. Venugopalan  TPI-MINN-94-7-T, Feb
1994; A. Ayala, J. Jalilian-Marian, L. McLerran, R.
Venugopalan, TPI-MINN-94-40-T, Jan 1995.

\bibitem{int} The Fall 1996 program of the  Institute for Nuclear Physics
in Seattle, WA will concentrate on this topic; contact organizers L.McLerran
or me.

\end{thebibliography}
\end{document}